\documentclass[aps,prb,twocolumn,superscriptaddress]{revtex4-2}
\usepackage{bm}
\usepackage{bbm}
\usepackage{hyperref}
\usepackage{graphicx}
\usepackage{natbib}
\usepackage{amsmath}

\begin{document}

\title{Theory of bound magnetic polarons in cubic and uniaxial antiferromagnets}

\author{Dawid Bugajewski}
\email{d.bugajewski@student.uw.edu.pl}
\affiliation{Faculty of Physics, University of Warsaw, Pasteura $5$, PL-$02093$ Warsaw, Poland}

\author{Carmine Autieri}
\email{autieri@MagTop.ifpan.edu.pl}
\affiliation{International Research Centre MagTop, Institute of Physics, Polish Academy of Sciences, Aleja Lotnikow $32/46$, PL-$02668$ Warsaw, Poland}

\author{Tomasz Dietl}
\email{dietl@MagTop.ifpan.edu.pl}
\affiliation{International Research Centre MagTop, Institute of Physics, Polish Academy of Sciences, Aleja Lotnikow $32/46$, PL-$02668$ Warsaw, Poland}

\date{\today}

\begin{abstract}
Motivated by a recent debate about the origin of remanent magnetization and the corresponding anomalous Hall effect in antiferromagnets and altermagnets, a theory of bound magnetic polarons (BMPs) in anisotropic magnetic semiconductors is developed. The theory describes quantitatively the experimentally observed magnitude of excess magnetization and its dependence on the magnetic field in cubic antiferromagnetic EuTe. In contrast to the cubic case, our theory predicts the presence of magnetization hysteresis below N\'eel temperature in antiferromagnets with uniaxial anisotropy. We show, employing material parameters implied by experimental and {\em ab initio} results, that the magnitudes of remanent magnetization and the coercive field are in accord with recent experimental observations for altermagnetic hexagonal MnTe.
While the altermagnets have an intrinsic contribution to the remanent magnetization and the anomalous Hall effect, our theory explains the origin of an extrinsic contribution. Our findings address, therefore, a question about the relative contribution to the remanent magnetization of bound magnetic polarons and weak ferromagnetism driven by the antisymmetric exchange interaction, the latter weakened by the formation of antiferromagnetic domains. Furthermore, we provide the theory of BMP spontaneous spin splitting, which can be probed optically.
\end{abstract}

\maketitle






Owing to strong sensitivity to doping, gating, and illumination on the one hand, and the presence of exchange split bands and generally high spin-ordering temperatures on the other, altermagnetic semiconductors \cite{Smejkal:2022_PRX,Bai:2024_AFM} constitute an appealing alternative to ferromagnetic semiconductors \cite{Dietl:2014_RMP,Jungwirth:2014_RMP} in search for new spintronic functionalities and relevant materials systems. A lot of recent attention has been devoted to new aspects of free magnetic polarons in antiferromagnets \cite{Nagaev:2001_JETFL,Klosinski2020,Wrzosek2024,Bohrdt:2020_NJP,Nielsen2021,Nyhegn2023} and altermagnets \cite{Steward2023_PRB}, earlier studied in the context of copper oxide superconductors  \cite{Zhang:1988_PRB}. However, one of the pertinent questions concerns the possibility of controlling magnetic properties by changing the occupation of in-gap impurity of defect states.

In this Letter, we develop the theory of bound magnetic polarons (BMPs) for the case of antiferromagnetic and altermagnetic semiconductors containing singly-occupied impurity states, extending previous approaches \cite{Mauger:1984_PRL,Liu:1986_PRB,Liu:1988_PRB,Castro_2004_EuroPJB,Ogarkov:2006_PRB,Kagan_2008_JPC} in two ways. First, we consider an experimentally relevant low-crystal symmetry case, such as hexagonal MnTe, in which, as we find, the presence of BMPs can lead to hysteretic ferromagnetic-like properties. Second, by taking into account thermodynamic fluctuations of magnetization, we go beyond both the zero-temperature limit and the mean-field approximation, an essential step at non-zero temperatures due to the finite size of BMP. Furthermore, we are interested in the thermal equilibrium case, so that the excitonic BMP, examined in a series of experimental and theoretical works in the context of EuTe luminescence \cite{Henriques:2022_JAP}, is not discussed here. Our starting point is the previous theory of BMPs in a paramagnetic phase \cite{Dietl:1982_PRL,Dietl:1983_PRB,Heiman:1983_PRB},  which has successfully described spin dynamics of quantum-dot electrons immersed in a bath of nuclear spins \cite{Dietl:2015_PRB} and earlier optical, charge transport, and magnetic properties observed in dilute magnetic semiconductors \cite{Dietl:1994_HB}. Interestingly, the BMP case constitutes an experimental realization of the central spin problem \cite{Gaudin:1976_JdP}, a persistent challenge of quantum statistical physics \cite{DeNadai:2024_PRB}. However, it is worth noting in this context that the semiclassical approach, within which the central spin $s$ is treated quantum mechanically whereas the surrounding spins $S$ classically, is quantitatively accurate for a dozen or more spins $S$ even if $S = 1/2$ provided that the statistical sum is calculated exactly, i.e., without, for instance, a mean-field approximation \cite{Dietl:2015_PRB}.

We show that the theory describes the time-honored observations of excess magnetization and its dependence on the magnetic field in cubic antiferromagnetic EuTe below N\'eel temperature $T_{\text{N}}$ \cite{Oliveira:1972_PRB,Vitins:1975_PRB}. In order to apply our results to altermagnetic hexagonal MnTe, we determine exchange integrals $J$ accounting for spin splittings of bands from first-principle computations. Making use of $J$ values and taking other relevant parameters from experimental data \cite{Komatsubara:1963_Jpn,Zanmarchi:1968_PhilResRep,Szuszkiewicz:2005_PSSc}, we show that our theory describes properly the experimentally found  \cite{Kluczyk:2024_PRB} orientation and magnitude of remanent magnetization $M_{\text{r}}$ along the hexagonal $c$ axis and the observed value of the coercive field $H_{\text{c}}$, also available from Hall resistance studies \cite{Kluczyk:2024_PRB,Gonzalez:2023_PRL}. We conclude that spin and orbital magnetization expected in MnTe from the antisymmetric exchange interactions  \cite{Mazin:2024_PRB,Autieri:2025_PRB} either compete with the BMP contribution or are masked by the formation of domains with antiparallel in-plane N\'eel vectors,  which results in a cancellation of magnetization resulting from spin canting along the $c$ axis driven by the antisymmetric exchange interactions. Such a cancellation does not occur in the case of BMPs, as the orientation of BMPs' magnetization along the $c$-axis is independent of the in-plane N\'eel vector direction.

{\em{Theory of bound magnetic polarons in antiferromagnetic semiconductors.}} \label{sec2}
We assume that in the case of antiferromagnets and altermagnets there are no spontaneous spin splittings at the band extrema, so that, a non-zero value of magnetization is necessary to generate spin splitting of a carrier bound to an impurity or defect.  We adopt, therefore, the magnetization of localized spins $\vec{M}(\vec{r})$ as an order parameter. For an effective-mass carrier bound to a singly ionized impurity, the BMP free energy assumes the form $F_{\text{BMP}} = E_{\text{imp}} + F_p$, where
\begin{equation}
E_{\text{imp}} = \langle\Psi|\frac{\hat{p}^2}{2m^*}- \frac{e^2}{\epsilon r}|\Psi\rangle,
\end{equation}
whereas $F_p$ is given by a functional integral over magnetization profiles \cite{Dietl:1983_PRB},
\begin{equation}
F_p = -k_{\text{B}}T\ln\int\mathcal{D}\vec{M}(\vec{r})P_p[\vec{M}(\vec{r})]/P_S[\vec{M}(\vec{r})].
\label{eq:Fp}
\end{equation}
The probability distribution $P_S$ of the localized spins' magnetization is determined by the free energy functional ${\cal{F}}_S[\vec{M}(\vec{r})]$, whose form can be determined from the experimental dependence of magnetization on the magnetic field in the absence of BMPs \cite{Swierkowski:1988_ActaPhysPol}. The distribution $P_p$ takes into account the presence of a bound carrier with a spin $s = 1/2$, according to
\begin{eqnarray}
&&
\hspace*{-.5cm} P_p[\vec{M}(\vec{r})] =  \nonumber \\
&&
\hspace*{.2cm} \exp(-{\cal{F}}_S[\vec{M}(\vec{r})]/k_{\text{B}}T)2\cosh(\Delta[\vec{M}(\vec{r})]/2k_{\text{B}}T),
\end{eqnarray}
where the bound carrier spin splitting $\vec{\Delta}$ is given by,
\begin{eqnarray}
\vec{\Delta}[{\vec{M}(\vec{r})}]=\frac{1}{g\mu_B}\int d\vec{r}J\vec{M}(\vec{r})|\Psi(\vec{r})|^2d\vec{r}+g^*\mu_B\vec{H}.
\label{eq:Delta}
\end{eqnarray}
Here, $g$, $g^*$, $J$, $\vec{H}$, and $\Psi(\vec{r})$ denote the Land\'e factor of the localized spins and bound carriers, the $spd$-$df$ exchange integral, the external magnetic field, and a trial wave function to be determined by a variational method, respectively. If the polaron contribution to $F_p$ becomes comparable or larger than the carrier binding energy in the absence of exchange coupling, the probability of magnetization distribution  $P_S[{\vec{M}(\vec{r})}]$  has to be taken beyond the Gaussian approximation \cite{Swierkowski:1988_ActaPhysPol,Benoit:1993_pssb}.

As we develop here the BMP theory for antiferromagnets and altermagnets with axial symmetry along the crystal $c$-axis,  the relevant material parameters $t$, such as magnetic susceptibility $\chi$ for the multidomain case, effective mass $m^*$, carrier Land\'e factor $g^*$ and exchange integral $J$ assume a diagonal tensor form, $t_a$, $t_a$, $t_c$ (where $t=\chi,m^*,g^*,J$).  It is also convenient to introduce a tensor of bound carrier spin-splitting, in the absence of the BMP contribution,
generated by macroscopic magnetization $M_{0}^{(i)}(T,H_i)$ and the external magnetic field $H_i$  applied along the direction $i$ = $a$ or $c$,
\begin{equation}
\Delta_0^{(i)} = J_iM_0^{(i)}(T,H_i)/g\mu_{\text{B}} + g^*_i\mu_{\text{B}}H_i,
\end{equation}
where in antiferromagnets at $T< T_{\text{N}}$, typically, $M_0^{(i)}(T,H_i) = \chi_i(T)H_i$ up to reaching a saturation value $M_{\text{SAT}}$. We also introduce the tensor of the BMP energy,
\begin{equation}
\varepsilon_p^{(i)} = \frac{J_i^2\chi_i}{4g^2\mu^2_{\text{B}}}\int d\vec{r}|\Psi(\vec{r})|^4,
\end{equation}
where $\chi_i = \partial M_i/\partial H_i$.  Making use of the identity,
\begin{equation}
\cosh x  = \frac{d}{d\lambda_0}\frac{1}{4\pi\lambda_0}\int d^3\lambda\delta(\lambda-\lambda_0) \exp\left(\vec{\lambda}\cdot\vec{x}\right)|_{\lambda_0=1},
\end{equation}
we obtain
\begin{equation}\label{fptheta}
F_p = -k_{\text{B}}T\ln\left(\frac{1}{2\pi}\left[\Theta'(1)-\Theta(1)\right] \right),
\end{equation}
where, introducing $b(\vec{r})=J|\Psi(\vec{r})|^2/2g\mu_{\text{B}}$,
\begin{eqnarray}\label{theta}
&&
\hspace{-.5cm} \Theta(\lambda_0)=\nonumber\\&&\int\limits_{\lambda=\lambda_0}d^2\lambda \exp\left[\frac{1}{k_{\text{B}}T}\int d^3r(F_S(0)-F_S(\vec{\lambda}b(\vec{r})))\right].
\end{eqnarray}
Here,  $F_S(\vec{H})$ is the localized spins' free energy in the magnetic field $\vec{H}$,  which can be obtained by inverting the experimental dependence $\vec{M}_0(T,\vec{H})$ into $\vec{h}(\vec{M})$ according to \cite{Swierkowski:1988_ActaPhysPol},
\begin{equation}
F(\vec{M}_0) = \int_0^{\vec{M}_0} d\vec{M} \vec{h}(\vec{M}) - \vec{H}\vec{M}_0,
\end{equation}
where $\vec{M}_0 = \vec{M}_0(\vec{H})$. Now, in the linear approximation $-\nabla_{\vec{H}}F_S(\vec{H})=\hat{\chi}\vec{H}$, we arrive at,
\begin{align}
F_p(\vec{H}=0) =
& -\frac{1}{2} \varepsilon_p^{(a)}
- k_{\text{B}} T \ln \bigg\{
2 \exp\left( \frac{\delta\varepsilon_p}{2k_{\text{B}}T} \right) \notag \\
& + \frac{\varepsilon_p^{(a)}}{k_{\text{B}}T}
\sqrt{ \frac{2\pi k_{\text{B}}T}{\delta\varepsilon_p} } \,
\text{erfi} \left( \sqrt{ \frac{\delta\varepsilon_p}{2k_{\text{B}}T} } \right)
\bigg\},
\end{align}
where $\delta \varepsilon_p  = \varepsilon_p^c - \varepsilon_p^a$.
By performing analogous calculations, it can be shown that
\begin{eqnarray}\label{hparez}
&&\hspace{1cm} F_p(\vec{H}\|\vec{c})=-\frac12 \varepsilon_p^a + \frac{\Delta_0^{c\,2}}{8\delta\varepsilon_p} \nonumber\\
&&-k_{\text{B}}T\ln\{\left[1+\exp(\Delta_0^{c}/k_{\text{B}}T)\right]\exp\left[\frac{(\Delta_0^{c}-2\delta\varepsilon_p)^2}{8k_{\text{B}}T\delta\varepsilon_p}\right] \nonumber\\
&&-\frac{\varepsilon_p^{a}}{2k_{\text{B}}T}\sqrt{\frac{2\pi k_{\text{B}}T}{\delta\varepsilon_p}}\left[\mbox{erfi}(\xi_-)-\mbox{erfi}(\xi_+)\right]\},
\end{eqnarray}
where $\xi_{\pm}=(\Delta_0^{c}\pm2\delta\varepsilon_p)/2\sqrt{2k_{\text{B}}T\delta\varepsilon_p}$ and $F_p(\vec{H}\perp \vec{c})$ is given by formula (\ref{fptheta}), where 
\begin{eqnarray}
&&
\hspace*{-.1cm}\Theta(\lambda_0)=2\pi\lambda_0^2\exp\left(\frac{\varepsilon_p^{a}}{2k_{\text{B}}T}\lambda_0^2\right) \nonumber\\
&&
\hspace{.4cm}\times\int\limits_{-1}^1dt\,\exp\left(\frac{\delta\varepsilon_p\lambda_0^2t^2}{2k_{\text{B}}T}\right)I_0\left(\frac{\lambda_0\Delta_0^{a}\sqrt{1-t^2}}{2k_{\text{B}}T}\right)
\end{eqnarray}
where $I_0$ is the $0$th order modified Bessel function. We have checked that for $\delta\varepsilon_p =0$, we recover the previously found form \cite{Dietl:1982_PRL},
\begin{equation}
F_p = -\frac{\varepsilon_p}{2} - k_{\text{B}}T\ln\left[2\cosh\left(\frac{\Delta_0}{2k_{\text{B}}T}\right) + \frac{4\varepsilon_p}{\Delta_0}\sinh\left(\frac{\Delta_0}{2k_{\text{B}}T}\right)\right].
\end{equation}
As trial functions we take $\Psi(\vec{r})=(\pi c^*a^{*2})^{-1/2}\exp(-\sqrt{(\rho/a^*)^2+(z/c^*)^2})$, where $a^*$ and $c^*$ are the bound carrier radii in the $c$-plane and along the $c$-axis, respectively. We also assume an anisotropic effective mass tensor and a spherically symmetrical impurity potential, as we assume the anisotropy of the dielectric constant is relatively small based on statistical data for various hexagonal crystals \cite{Lou:2025_Faraday}. In order to obtain the localization radii, the total BMP free energy, $E_{\text{imp}} + F_p$ is minimized with respect to $a^*$ and $c^*$.

It should be noted that the results above assume a linear dependence $\vec{M}(\vec{H})$, which is valid provided that $|\vec{M}|< M_{\mathrm{SAT}}=Sg\mu_BN_S$ when calculating $\Theta$ from Eq.~(\ref{theta}). This inequality may, however, be violated -- but only for Bohr radii $a^*,c^*$ satisfying
\begin{eqnarray}\label{condition}
c^*a^{*2}\leq \frac{\max(J_c\chi_c,J_a\chi_a)}{2\pi g\mu_0\mu_B M_{\mathrm{SAT}}}.
\end{eqnarray}
If the susceptibilities $\chi_i$ are small enough (like for MnTe), the condition (\ref{condition}) is not satisfied for any $a^*,c^*$ of realistic magnitudes; one can therefore perform calculations by neglecting the existence of a magnetization upper bound $M_{\mathrm{SAT}}\rightarrow\infty$, and numerically find the correct localization radii. Otherwise,  formulas for $F_p$ need to be modified by, e.g., setting $\vec{M}=M_{\mathrm{SAT}}\vec{H}/|\vec{H}|$ for $|\hat{\chi}\vec{H}|> M_{\mathrm{SAT}}$. This approach will be used in numerical calculations for EuTe.

The determined free energy  $F_p(T,H)$ provides directly the BMP contribution to the specific heat and magnetization \cite{Dietl:1982_PRL,Dietl:1983_PRB,Wojtowicz:1993_PRL} as well as to the activation energy of conductivity \cite{Jaroszynski:1985_SSC}. In view of anticipated experiments on spin-flip Raman scattering \cite{Heiman:1983_PRB,Nawrocki:1981_PRL} and optical absorption \cite{Dobrowolska:1984_PRB}, we provide the expected probability distribution of bound carrier spin splitting,
 \begin{eqnarray}
 P_p(\vec{\Delta}) = Z^{-1}2\cosh(\Delta/2k_{\text{B}}T) \nonumber\\
 \times\exp\left[-\frac{(\Delta_c - \Delta_0^{c})^2}{8k_{\text{B}}T\varepsilon_p^{c}} - \frac{(\vec{\Delta}_a -\vec{\Delta}_0^{a})^2}{8k_{\text{B}}T\varepsilon_p^{a}}\right],
\end{eqnarray}
where $Z$ is a normalization constant and $\Delta = (\Delta_c^2 + \vec{\Delta}_a^2)^{1/2}$.
In the experimentally relevant case, $\Delta \gg k_{\text{B}}T$ and $\varepsilon_p^{c} \gg \varepsilon_p^{a}$ the most probable value of the spin splitting $\bar{\Delta} = \Delta_0^{c} + 2\varepsilon_p^{c}$. Under these conditions, $F_p = -\Delta_0^{c}/2 - \varepsilon_p^{c}/2$, as the formation of magnetic polaron lowers both the energy and entropy.

\begin{figure}
\includegraphics[width=.95\columnwidth]{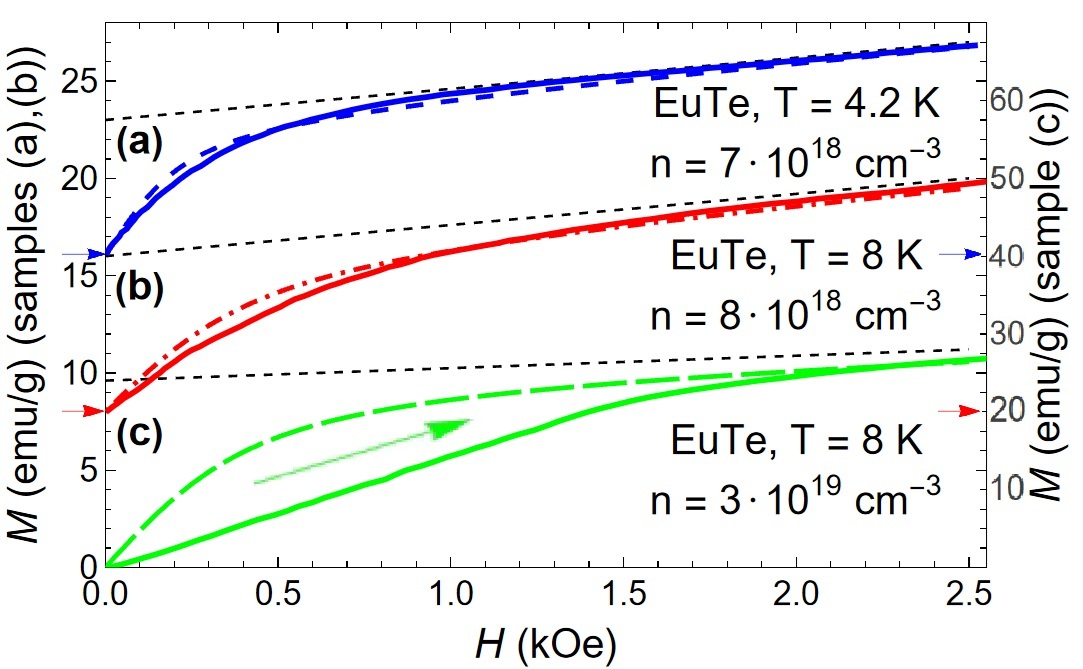}
\caption{\label{figEuTe} Dependence of the magnetic moment per gram $M$ on the magnetic field calculated from Eq.~\ref{EuTe} (dashed curves) compared to experimental results (solid curves) at (a) 4.2\,K \cite{Oliveira:1972_PRB} and (b),(c) 8\,K \cite{Vitins:1975_PRB}. The fitted BMP concentrations $n$ are display in the legend.  The straight dotted lines show a linear asymptotic behavior and provide the magnitude of magnetic susceptibility $\chi$. An offset $16$ and $8\,$emu/g was added to (a),(b) respectively for readability - two arrows next to the vertical axis indicate the positions of $M=0$ for those two plots.}
\end{figure}

{\em{Comparison to experimental results for EuTe and MnTe.}}\label{sec3}
We first present results of numerical calculations for cubic antiferromagnetic EuTe without any long-range order in the N\'eel vector orientation, for which $\hat{\chi}$ and $J$ are reduced to scalars; so that $a^*=c^*$ has been assumed in the family of trial functions. However, a rather large magnitude of magnetic susceptibility, resulting from a low value of $T_{\text{N}} = 9.6$\,K, makes the magnetization generated by BMP attain locally the saturation value $M_{\mathrm{SAT}}$, which has to be incorporated into the form of the functional ${\cal{F}}_S[\vec{M}(\vec{r})]$, as already mentioned. A consensus has emerged that the conduction band in EuO and Eu monochalcogenides is built of Eu $d$ orbitals for which the intra-atomic potential exchange   $J_{5d4f} = 215$\,meV \cite{Russell:1941_PR}. This atomic value compares favorably with the observed and theoretical magnitudes of exchange band splitting in EuO $J = 170$ \cite{Steeneken:2002_PRL} and 230\,meV \cite{Tong:2014_PRB}, respectively. We adopt, therefore, $J=0.2$\,eV together with   $m^*=0.3m_e$, $\epsilon=6.9$, $N_0=4/a^3=1.4\times10^{22}\,$cm$^{-3}$ (like in \cite{Gratens:2020_APL}), $\chi=1.61\times10^{-3}\,\mathrm{emu}/\mathrm{gOe}$ (slope of $M(H)$ for samples without BMPs \cite{Oliveira:1972_PRB,Vitins:1975_PRB}), and  $g^* = g =2.0$, the latter expected for both Eu$^{+2}$ and Mn$^{+2}$ in the orbital singlet configuration. For these parameters, we determine magnetization according to,
\begin{eqnarray}\label{EuTe}
M(H)=-n\partial F_p/\partial H + \chi H,
\end{eqnarray}
where $n$ is the bound carrier concentration that, in general, may depend on temperature, and $F_p$ is calculated for $a^*$ resulting from minimization of $F_p$ with respect to $a^*$.

Figure~\ref{figEuTe} presents a comparison of our theoretical results to experimental data for EuTe below N\'eel temperature \cite{Oliveira:1972_PRB,Vitins:1975_PRB}. As cubic anisotropy is typically small, even allowing for a larger magnitude of magnetic susceptibility in the plane perpendicular to the local N\'eel vector, magnetization brought about by BMPs should show no hysteresis and vanish in the absence of a magnetic field. However, in $H>0$, BMPs' magnetic moments align along the magnetic field, and the BMP contribution shows up. Importantly, this additional magnetization was observed neither for metallic samples \cite{Shapira:1972_PRB} nor for a sample with an unmeasurably low electron concentration $n_e$ at 300\,K \cite{Vitins:1975_PRB}, which supports its interpretation in terms of BMPs. The agreement between theory and experiment is quite good for samples with donor densities below $10^{19}$ cm$^{-3}$ [plots (a),(b)]. For $n_e = 4\times10^{18}$\,cm$^{-3}$ \cite{Vitins:1975_PRB}, we find $n = 8\times10^{18}$\,cm$^{-3}$, as in this strongly localized case room temperature electron concentration is presumably smaller than the bound carrier density at helium temperatures. In contrast, the agreement is poor for the sample with the highest value of $n_e =6\times10^{19}$\,cm$^{-3}$ \cite{Vitins:1975_PRB} [plot (c)], probably due to interactions between BMPs. The obtained effective Bohr radius $a^*=0.71\,$nm leads then to $n_e^{1/3}a^*=0.28$, which somewhat surpasses the Mott critical value $0.26$ marking the delocalization transition. Accordingly, the low-temperature BMP concentration determined from magnetization data in Fig.~\ref{figEuTe} is smaller, $n = 3\times10^{19}$\,cm$^{-3}$, as only bound carriers with energies below the mobility edge form BMPs.

\begin{figure}
\includegraphics[width=.8\columnwidth,angle=270]{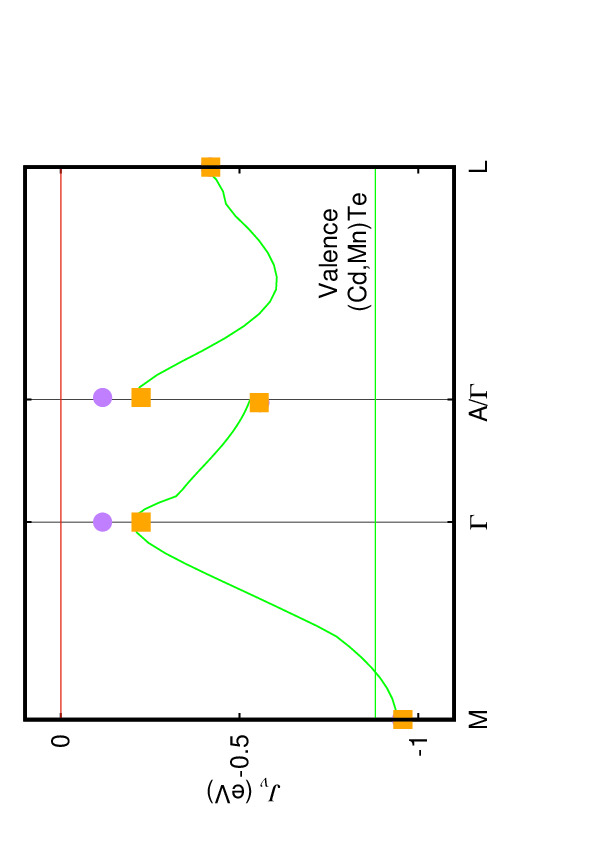}
\vspace{-10mm}
\caption{\label{fig:ab initio} Computed exchange energy $J_v$ for the valence band of MnTe in the
non-relativistic case (green lines) and in the
 relativistic case with saturated magnetization along the $a$-axis
(purple dots) and along the $c$-axis (orange triangles). The
horizontal line is the experimental value of $J_v$ at the $\Gamma$ point of the valence band in (Cd,Mn)Te (see, Ref.\,\onlinecite{Autieri:2021_PRB}).}
\end{figure}

We now turn to the numerical analysis of BMP magnetic properties in bulk hexagonal $p$-type MnTe. We assume  $g^* = g = 2.0$ and take the values of dielectric constant, lattice parameters, magnetic susceptibility, and effective mass tensor from Refs.~\onlinecite{Oleszkiewicz:1988_TSF,Szuszkiewicz:2005_PSSc,Kluczyk:2024_PRB,Zanmarchi:1968_PhilResRep}. The $J$ tensor has been obtained by a method applied previously for (Cd,Mn)Te and (Hg,Mn)Te \cite{Autieri:2021_PRB} with the implementation of the density functional theory described elsewhere \cite{Chen:2025_npj}. In agreement with previous results \cite{Faria:2023_PRB}, we find the valence band maximum at the A point of the Brillouin zone and determine the magnitude of the band splitting for the ferromagnetic arrangement of Mn spins. As shown in Fig.~\ref{fig:ab initio}, exchange splitting of the topmost valence band at the A point corresponds to $J_a = J_c = -0.5$\,eV. This time, since the magnetic susceptibilities along and perpendicularly to the $c$-axis are significantly smaller than $\chi_{\mathrm{EuTe}}$, there is no need to allow for a finite value of $M_{\mathrm{SAT}}$.

Under thermal equilibrium and in $H=0$, BMP average magnetization vanishes due to symmetry, while for $H\rightarrow\infty$ BMP moments are aligned along the magnetic field. However, uniaxial anisotropy of magnetic susceptibility at $T< T_{\text{N}}$ can result in a hysteretic behavior of BMP magnetization when cycling the external magnetic field. According to the Stoner-Wolfarth model, the coercivity field $H_c =2K/m$, where the anisotropy energy $K$ is given by a difference between the BMP free energy for the BMP magnetic moment $m$ aligned along the hard and easy axis,
\[
K= \lim\limits_{H\rightarrow\infty}[F_p(\vec{H}\perp \vec{c})-F_p(\vec{H}||\vec{c})].
\]
The BMP magnetic moment is  determined from Eq.\,(\ref{hparez}) as,
\begin{eqnarray}
&&
\hspace*{-1cm}m=-\partial F_p/\partial H |_{{H}\rightarrow\infty}=\nonumber\\&&\hspace*{3cm}\frac12|g^*_c\mu_{\text{B}}+ J_{c}\chi_{c}/g\mu_{\text{B}}|,
\end{eqnarray}
so that, within our model, the remanent magnetization for the magnetic field cycled along the $c$ axis is expected to be determined by $M_{\mathrm{REM}} = nm$.

\begin{figure}[h!]
\vskip .2cm
\includegraphics[width=.95\columnwidth]{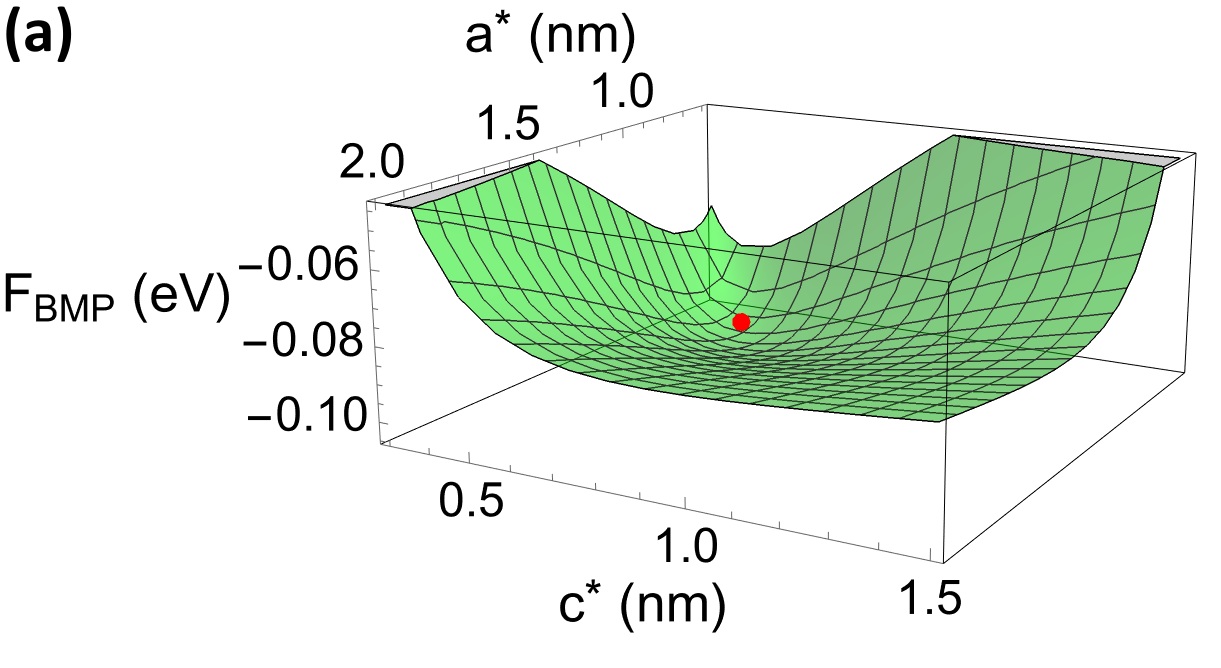}
\vskip .2cm
\includegraphics[width=.95\columnwidth]{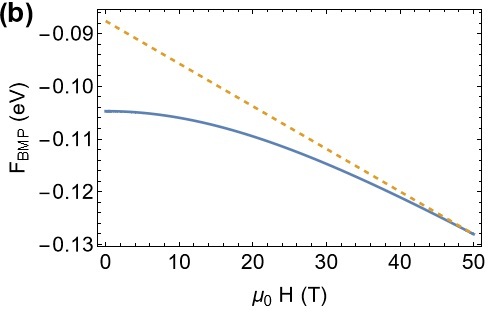}
\caption{\label{mnte} (a) dependence of BMP free energy $F_p$ on the effective Bohr radii along the $c$-axis ($c^*$) and in the $c$-plane ($a^*$). The red point depicts a minimum found by the variational procedure. (b) dependence of the BMP free energy (minimized with respect to radii $a^*$ and $c^*$) on the  magnetic field $H$ applied along the $c$-axis. The straight line is a linear asymptote corresponding to BMP's magnetic moment parallel to the magnetic field. Temperature $T=250\,$K for both plots.}
\end{figure}

Figure \ref{mnte} presents the numerical implementation of the BMP theory for bulk MnTe at experimental temperature $T=250\,$K \cite{Kluczyk:2024_PRB} -- the dependence of BMP free energy in $H=0$ on the effective localization radii $a^*,c^*$ and the dependence of $F_p$ minimized with respect to $a^*,c^*$ on $\vec{H}$. For instance, $a^*=0.89\,$nm, $c^*=0.54\,$nm for $H=0$; these values depend rather weakly on $H$ in the considered range, but it should be noted that they are quite sensitive to the values of $m^*$ and $\epsilon$. These values and the Mott condition $(na^{*\,2}c^*)^{1/3}=0.26$ imply that the delocalization transition occurs at $n = 4\times10^{19}$\,cm$^{-3}$.

Carrier localization at low temperatures and the presence of the anomalous Hall effect make the determination of the bound carrier concentration difficult. The experimental value $M_{\mathrm{REM}}\sim 5\times10^{-5}\,\mu_{\text{B}}$/Mn \cite{Kluczyk:2024_PRB} implies, within our model,  $n \simeq 10^{16}-10^{17}$\,cm$^{-3}$ at 250\,K. This evaluation constitutes an upper limit, as one expects that magnetization  brought about by weak ferromagnetism specific to MnTe will align with the BMP magnetic moment. Obviously, studies of remanent magnetization as a function of bound carrier concentration would allow to verify the role of BMPs.
Within our model, we also obtain  $K=0.9\,$meV and $H_c=21\,$kOe at that temperature, which appears to compare favorably with the experimental value $H_c = 25 \pm5$\,kOe \cite{Kluczyk:2024_PRB,Gonzalez:2023_PRL} but leads to rather low superparamagnetic blocking temperature $T_\text{B}$. Neglecting coupling between BMPs, $K > 0.4$\,eV  would lead to  $T_\text{B} > 250$\,K. Actually, strong anisotropy of weak ferromagnetism magnetization \cite{Chen:2025_npj} and the exchange bias by surrounding antiferromagnetic lattice \cite{Castro_2004_EuroPJB,Skurmyev:2003_Nature} can enhance the magnitude of $K$ considerably.  In particular, the exchange coupling causes a distortion of the antiferromagnetic spin lattice in the range of the order of $(|J|/K_c)^{1/6}$ lattice sites \cite{Castro_2004_EuroPJB}, where $J$, $K_c$ are exchange integrals between neighboring Mn ions and crystalline anisotropy, respectively. For $|J|/K_c\sim 2500$, as in the spin canting modelling  \cite{Kluczyk:2024_PRB}, we find that the distortion extends over 3-4 lattice sites, i.e., involves $\sim 600-700 $ Mn ions, to be compared to around $40$ spins in the BMP volume $4\pi a^{*2}c^*/3$. The resulting coupling between BMPs will enhance the $K$ value even more.
 
{\em{Conclusions and outlook.}}\label{sec4}
We have introduced a new ingredient into the physics of altermagnetic semiconductors, namely the bound magnetic polarons. In the case of single-domain microstructures, BMP magnetization will compete with intrinsic weak ferromagnetism \cite{Chen:2025_npj} allowed for some altermagnets in the presence of a spin-orbit interaction \cite{cheong2025altermagnetismclassification}. In the multidomain situation, non-zero magnetization resulting from weak ferromagnetism cancels out, leaving the BMP contribution to dominate, most probably the case of MnTe samples discussed here, taking into account the results of nanoscale magnetization imaging of MnTe \cite{Amin:2024_N}. However, the presence of local weak magnetization will lead to an additional shift or broadening of the BMP spin-resonance line. Hence, spin-flip Raman scattering and related techniques could constitute a valuable tool for probing magnetization distribution and dynamics of altermagnetic semiconductors at the nanoscale. We also note that the presence of large bound carrier spin-splitting in Raman scattering, but significantly smaller magnetization than predicted here, will imply that we deal with a compensated magnet \cite{Finley:2020_APL} or an altermagnet allowing for zero-field exchange splitting at the relevant band extremum. 

\section*{Acknowledgements}
This research was supported by the “MagTop” project (FENG.02.01-IP.05-0028/23) carried out within the “International Research Agendas” programme of the Foundation for Polish Science co-financed by the European Union under the European Funds for Smart Economy 2021-2027 (FENG). We further acknowledge access to the computing facilities of the Interdisciplinary Center of Modeling at the University of Warsaw, Grant g91-1418, g91-1419, g96-1808 and g96-1809 for the availability of high-performance computing resources and support.

\section*{Data availability}
The data that support the findings of this article are openly available \cite{data}.

\end{document}